\documentclass{aa}
\usepackage{txfonts}
\usepackage{graphicx}

\newcommand{\lsim}{\raisebox{-.4ex}{$\stackrel{<}{\scriptstyle \sim}$}}
\newcommand{\msim}{\raisebox{-.4ex}{$\stackrel{>}{\scriptstyle \sim}$}}

\def \hii {H~{\sc ii}}
\def \heii {He~{\sc ii}}
\def \eso   {ESO\,338-4}
\def \zw   {I\,Zw\,18}

\def\change{ }
\newcommand{\rev}{  }
\def\refrep{}

\def\ergs{erg s$^{-1}$}

\def\msun{\ifmmode M_{\odot} \else M$_{\odot}$\fi}
\def\msunyr{\ifmmode M_{\odot} {\rm yr}^{-1} \else M$_{\odot}$ yr$^{-1}$\fi}
\def\zsun{\ifmmode Z_{\odot} \else Z$_{\odot}$\fi}
\def\lsun{\ifmmode L_{\odot} \else L$_{\odot}$\fi}

\begin{document}

\title{Ionization of \heii\  in star-forming galaxies by X-rays from 
cluster winds and  superbubbles}
\titlerunning{\heii\ ionization in starburst galaxies by cluster winds and superbubbles}
\authorrunning{L.\,M.~Oskinova \& D.~Schaerer}
\author{Lidia M. Oskinova\inst{1}
\and Daniel Schaerer \inst{2,3}
}

\institute{
Institute for Physics and Astronomy, University Potsdam, D-14476 Potsdam,
Germany
\and
Observatoire de Gen\`eve, D\'epartement d'Astronomie, Universit\'e de Gen\`eve, 51 Chemin Pegasi, 1290 Versoix, Switzerland
\and
CNRS, IRAP, 14 Avenue E. Belin, 31400 Toulouse, France}

\date{Received <date> / Accepted <date>}

\abstract{The nature of the sources powering nebular \heii\  emission in star-forming 
galaxies remains debated, and various types of objects have been considered, including 
Wolf-Rayet stars, X-ray binaries,  and Population\,III stars. Modern X-ray observations 
show the ubiquitous presence of hot gas filling star-forming galaxies.
We use a collisional ionization plasma code to compute the specific \heii\ ionizing flux 
produced by hot gas and show that if its temperature is not too high  ($\protect\la 2.5$\,MK), 
then the observed levels of soft diffuse X-ray radiation  could  explain \heii\ 
ionization in galaxies.   
To gain a physical understanding of this result, we propose a  model that combines 
the hydrodynamics of cluster winds and hot superbubbles with observed populations of 
young massive clusters in galaxies. We find that in low-metallicity galaxies, the temperature 
of hot gas is lower and the production rate of \heii\ ionizing  photons is higher compared 
to high-metallicity galaxies. {\change The reason is that the  slower
stellar winds of massive stars in 
lower-metallicity galaxies input less mechanical energy in the ambient medium. Furthermore, we show 
that ensembles of star clusters up to $\sim 10-20$ Myr old in galaxies}
can produce enough soft X-rays to 
{\refrep induce} nebular \heii\ emission. 
We discuss observations of the template low-metallicity galaxy \zw\ 
and suggest that the \heii\ nebula in this galaxy is powered by a hot superbubble. Finally, 
appreciating the complex nature of stellar feedback, we suggest that  soft X-rays from hot superbubbles 
are among the dominant sources of \heii\ {\refrep ionizing flux} in low-metallicity star-forming galaxies. }


\maketitle

\section{Introduction}

A narrow nebular \heii\ $\lambda 4686$\,\AA\ emission line is observed in many
low-metallicity (low-$Z$)  star-forming galaxies, implying the presence of sources emitting 
copious amount of He$^+$ ionizing photons, which are not expected from normal stellar populations.
\citep[e.g.,][]{Guseva2000,Shirazi2012}
The nature of these sources remains mysterious,  and different proposed explanations are under debate. 


Hot Wolf-Rayet (WR) stars, stars stripped in binary interactions, or binary {\refrep systems 
in general} have been suggested as possible sources of He$^+$ ionizing photons
\citep[e.g.,][]{Schaerer1996,Dionne2006,Szecsi2015,Gotberg2019,Sander2020,Doughty2021}.
Other authors have proposed the presence of very metal-poor or supermassive stars 
\citep{Cassata2013,Kehrig2015},
radiative shocks
\citep[][]{Thuan2005,Izotov2012,Plat_2019}, 
or other mechanisms
\citep{Barrow2019,PerezMontero2020} to explain nebular \heii\ emission at low metallicity.

Recently, high-mass X-ray binaries (HMXBs) that consist of  a neutron star or a black hole
accreting matter lost by a  massive star donor have received special attention.
Among them, the ultraluminous 
X-ray sources (ULXs) have {\refrep the} highest X-ray  luminosities (which we define throughout this paper to be in the 
0.3--10.0\,keV energy range), $L_{\rm X}\msim 10^{39}$\,erg\,s$^{-1}$. \heii\ and He\,{\sc iii} nebulae 
 are commonly observed around ULXs \citep[e.g.,][]{Pakull2002, Pakull2003,Moon2011Large-Highly-Io,Binder2018}. 

Among galaxies with similar star-formation rates (SFRs), 
{\refrep 
a higher X-ray output is produced by HMXBs in galaxies with lower metallicities 
\citep{Mineo2012, Lehmer2021}.}
Noting this trend and the similarity of the metallicity dependence of the \heii$/$H$\beta$  line intensity, 
\citet{Schaerer2019} suggested that \heii\ nebulae in star-forming galaxies are ionized by ULXs.
However, the  recognition of ULXs as prime ionizing sources in \heii\ emitting galaxies is debated.
For example, \citet{Kehrig2015, Kehrig2021} argue that the ULX radiation  
is insufficient to explain the observed \heii\ emission in the well-studied low-$Z$ galaxy \zw\ \citep[although new integral field spectroscopic observations provide a less definite answer on this issue; see][]{Rickards2021}.
Also, \citet{Oskinova2019} show that the contribution of  ULXs to the \heii\ ionization in the starburst galaxy \eso\ is small.  
From observations of 18 high-redshift galaxies, \citet{Saxena2020} conclude 
{\refrep that HMXBs  and weak (or obscured) active galactic nuclei} are unlikely to be the dominant 
producers of He$^+$ ionizing photons in these galaxies. 

{\refrep Finally, using simple photoionization models that combine HMXBs described by a multicolor disk model with a normal stellar population, \cite{Senchyna2020} have concluded that HMXBs are inefficient \heii\ ionizing photon producers.
However, \cite{Simmonds2021} challenge this conclusion on an empirical basis and using photoionization models, pointing out that the ionizing flux strongly depends on the poorly known shape of the spectral energy distribution (SED) of the ULXs. Clearly, the impact of compact X-ray sources on the interstellar medium (ISM) in galaxies needs further studies. In addition, other sources and emission processes may also be of importance.}

In this article we {\refrep introduce} a model that explains \heii\ emission in 
star-forming galaxies via  the radiative output of star cluster winds 
and associated superbubbles. The paper is organized as follows. The observations of 
hot diffuse gas in galaxies are reviewed in Sect.~\ref{sec:obs}. The specific 
\heii\ ionizing flux produced by hot plasma is calculated in  Sect.~\ref{sec:q}.
The parameterizations of the temperature, the luminosity of cluster winds and superbubbles,  
and the \heii\ ionization parameter are presented in Sect.~\ref{sec:sb}. The  
populations of star clusters are incorporated in the model in Sect.~\ref{sec:gal}.
Section \ref{sec:zw} provides a comparison of the model predictions with observations. 
The summary of this work is given in Sect.~\ref{sec:sum}.

\section{Empirical scaling between the X-ray luminosity of hot diffuse gas and the star-formation rate 
in galaxies} 
\label{sec:obs}

Young massive star clusters drive  winds and blow superbubbles filled by hot   gas
\citep{MacLow1988,Strickland2004, Silich2005,Keller2014,Kavanagh2020}. 
Fortunately, the subarcsecond angular resolution offered by the {\em Chandra}  
telescope allows X-ray point sources, localized areas of diffuse X-ray emission 
(e.g.,\ supernova remnants), and  hot extended superbubbles in star-forming galaxies to be disentangled.

A large sample containing 21 {\refrep local} galaxies with SFRs between $\sim 0.1$ and 
$\sim 20\,M_\odot$\,yr$^{-1}$ and a broad range of stellar masses was studied by \citet{Mineo2012}. 
Hot diffuse gas 
was detected in all sample galaxies. Their X-ray spectra have at least 
one soft  spectral component that could be modeled as originating from plasma with a temperature 
$\left<kT\right>\approx 0.24$\,keV ($\sim 2.8$\,MK). 
\citet{Mineo2012} derive the intrinsic luminosity of hot gas in their sample galaxies  using the {\sc mekal} 
plasma models \citep{mekal} as
$L_{\rm X}(0.3-10\,{\rm keV})/{\rm SFR}\approx 7 \times 10^{39}$ \ergs/(\msunyr).
{\refrep
An even larger sample of 49 galaxies with various morphological types was studied by  
\citet{Smith2018} and \citet{Smith2019}, who found that
$L_{\rm X}^{\rm gas}(0.3-8\,{\rm keV})/{\rm SFR}\approx 6 \times 10^{39}$ \ergs/(\msunyr).}

We take these empirical measurements as a guidance to approximate the scaling between 
the total X-ray luminosity of diffuse hot gas in a star-forming galaxy and its SFR 
as  
\begin{equation}
\log(L_X/{\rm SFR}) \approx 39.8 ~~~~[{\rm erg\,s}^{-1}/(M_\odot\,{\rm yr}^{-1})].
 \label{eq:sfr40}
\end{equation}

\section{\heii\ ionizing flux from hot thermal plasma}
\label{sec:q}
Following \citet{Schaerer2019}, {\refrep we  introduce a specific ionizing parameter, $q$ [erg$^{-1}$], that describes 
the \heii\ ionizing photon flux,  $Q({\rm He}^+)$ [s$^{-1}$], emitted 
by a hot plasma per X-ray luminosity, $L_{\rm X}$ [erg\,s$^{-1}$]:}
%
\begin{equation}
 q=Q({\rm He}^+)/L_{\rm X}.
\label{eq:q}
\end{equation}
In this paper we evaluate the \heii\ ionizing photon flux in the $0.054$--$0.5$\,keV range, 
while the X-ray luminosity is in $0.3$--$10.0$\,keV range unless otherwise noted. 

\begin{figure}[]
\centering
\includegraphics[width=0.99\columnwidth]{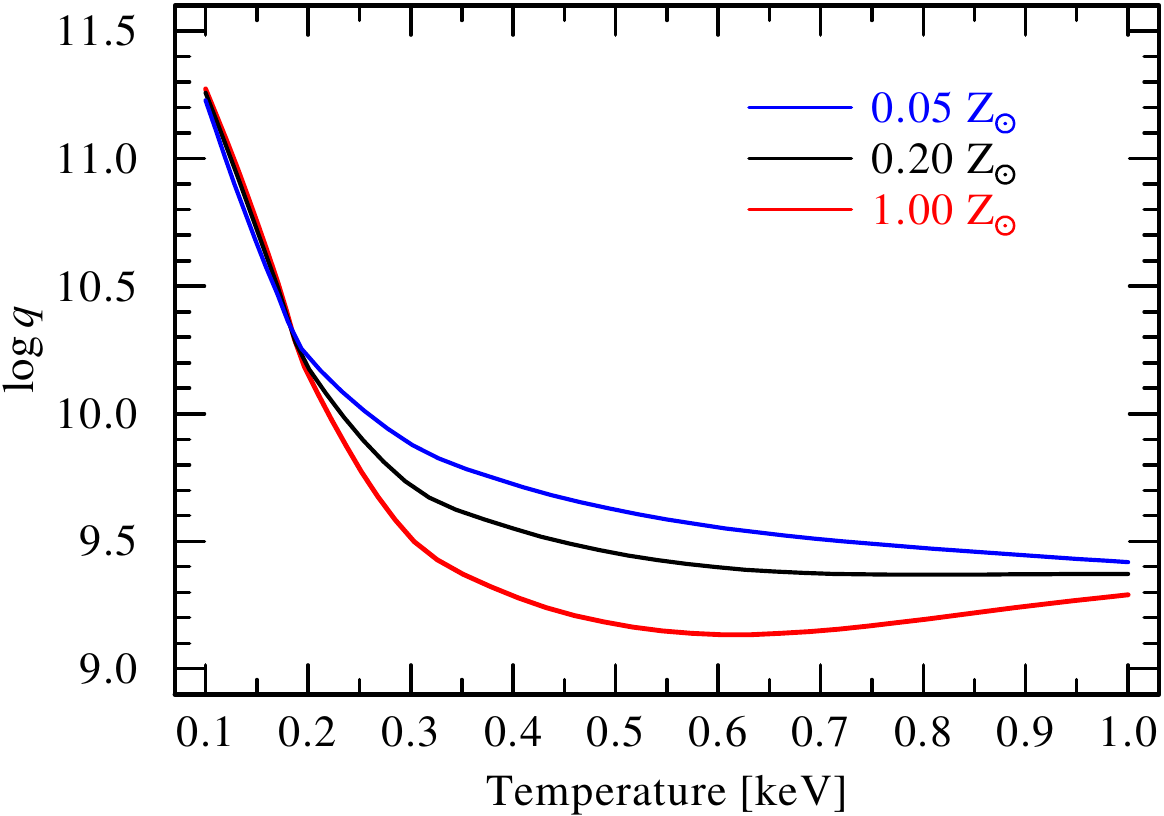}
\caption{Dependence of the parameter $q$ (see Eq.~\ref{eq:q}) on the 
plasma temperature for different metallicities, $Z$ (see legend in the upper-right corner) 
according to the collisional plasma model {\sc apec}.} 
\label{fig:qT}
\end{figure}

{\refrep To evaluate the ionizing flux produced by a hot thermal plasma, we computed  $q$ as a function of the plasma 
temperature, $T$, and metallicity, $Z$, by using the Astrophysical Plasma Emission Code  ({\sc apec})
 for collisionally ionized plasma in the optically thin limit}  \citep[][]{apec}  as implemented 
in the most recent editions of the X-ray spectral fitting package {\em Xspec} \citep{xspec1996}.
The results are shown in Fig.\,\ref{fig:qT}.  One can see that $q(T, Z)$ 
increases steeply  with decreasing temperature and becomes independent of $Z$ for  
$T$ $\lsim$ $0.2$\,keV. 

According to \citet[][and references therein]{Schaerer2019}, the ratio
between the \heii\,$\lambda 4686$ and H$\beta$ line intensities in star-forming galaxies is 
\begin{equation}
I(4686)/I({\rm H}\beta)\approx 2\times \frac{q \times L_{\rm X} }{10^{53}\times {\rm SFR}}.
\label{eq:ii1} 
\end{equation}
Substituting the empirical scaling relation Eq.~\ref{eq:sfr40} for $L_{\rm X}$ yields  
\begin{equation}
I(4686)/I({\rm H}\beta)\approx 
\frac{q \times 10^{39.8}\times {\rm SFR}}{10^{53}\times {\rm SFR}}=
6 \times 10^{-14}\times q(Z, T_{\rm X}).
\label{eq:ii} 
\end{equation}
Remarkably, the predicted ratio  is (nearly) independent of  SFR, which is\ in accordance with observations 
 \citep[][and references therein]{Schaerer2019}.

{\refrep 
As a next step, we 
made the implicit assumption that the metallicity of the hot gas is the same as the average galactic metallicity. The 
latter is often evaluated from nebular diagnostics \citep{Kewley2002} and,  therefore, characterizes the 
ISM in the vicinity of star-forming regions. Of course, stellar winds and multiple generations of supernovae (SNe) enrich 
the hot gas in the ISM. However, these effects are mild for bubbles around the ``middle age'' 
($\lsim 10$\,Myr old) clusters we consider here \citep[][]{Cheng2021}.

According to Fig.\,\ref{fig:qT}, at $Z\leq 0.2Z_\odot$ and for a broad range of temperatures 
between  0.1 and 1.0\,keV (1--11\,MK)  the parameter $q$ is in the range $9.5 < \log{q} < 11.5$. }
Including this in Eq.~(\ref{eq:ii}), we obtain  for such low-$Z$ galaxies 
\begin{equation}
I(4686)/I({\rm H}\beta) \approx 10^{-3.7}-10^{-1.7} 
\label{eq:iiq} 
\end{equation}
(i.e., in the observed range; see, e.g., Fig.\,1 in \citealt{Schaerer2019}).  This implies that the
hot plasma filling the ISM in low-$Z$ star-forming galaxies is an important source of \heii\ ionizing 
radiation. 


\section{Ionizing radiation from model cluster winds and superbubbles}
\label{sec:sb}

\begin{figure}[]
\centering
\includegraphics[width=0.9\columnwidth]{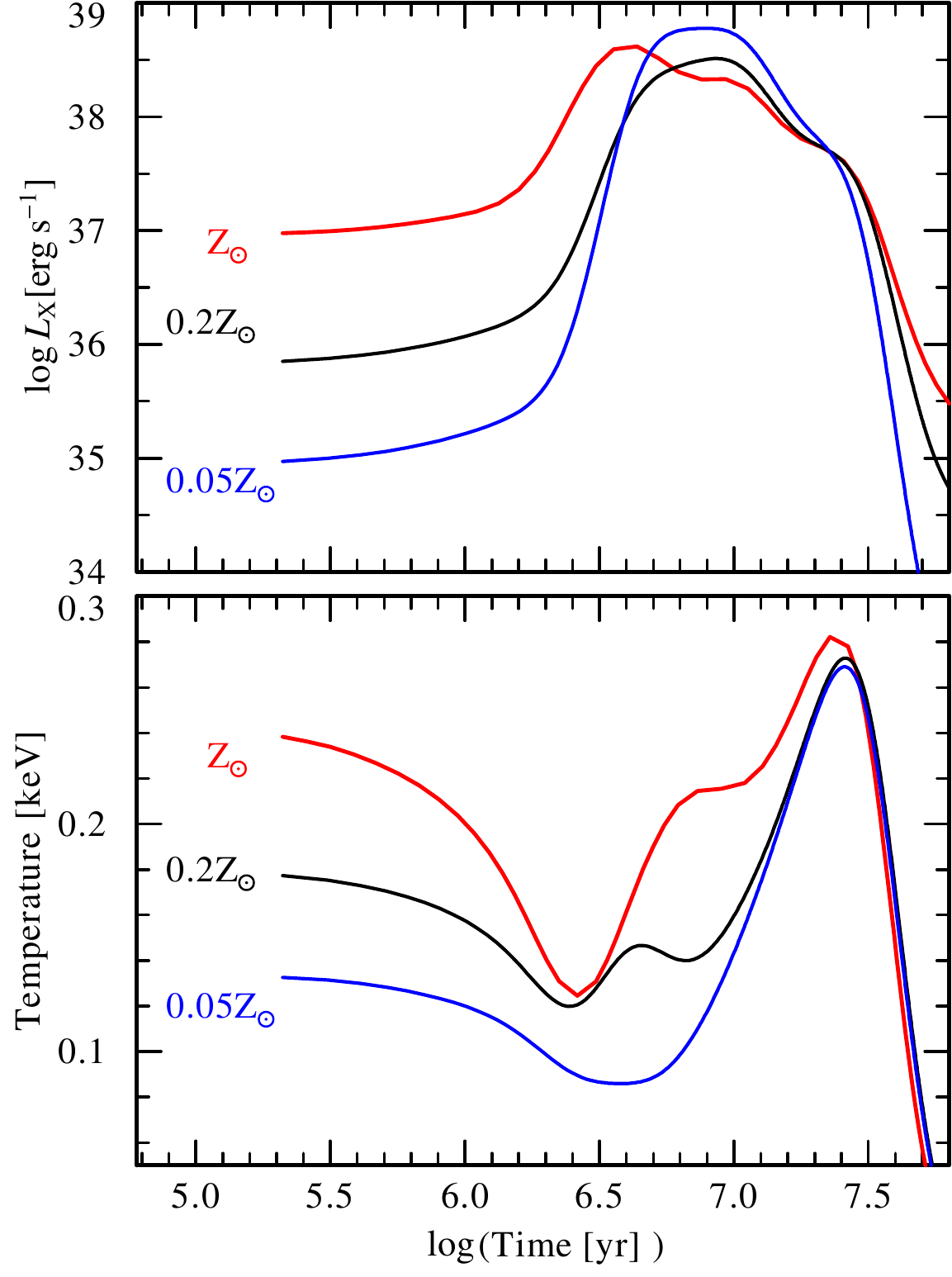}
\caption{X-rays from a model cluster of $10^6\,M_\odot$ at different metallicities, 
assuming an instantaneous burst of star formation and a standard (Salpeter) initial mass function 
\citep[computed according to Eqs.~(3\,--7) in ][ but adopting $\alpha=0.05$ and 
$\dot{M}_{\rm cool}=\dot{M}_\ast$; see  Sect.~\ref{sec:sb}]{Osk2005}.
{\it Upper panel}: Dependence of X-ray luminosity on cluster age. The curves are
smoothed to correct for the computational discontinuity of {\em Starburst99}
models  for SN rates \citep[see][]{SB99}. {\it Lower panel}:
Dependence of temperature on  cluster age at different metallicities.} 
\label{fig:txlx}
\end{figure}

{\refrep The estimates made in the previous section (Sect.\,\ref{sec:q}) show that 
the radiation from gas with temperatures 1--10\,MK supplies a significant amount of \heii\ ionizing photons. In 
star-forming galaxies such hot gas is found within superbubbles heated by the feedback 
from young massive star clusters. Below we consider how the 
feedback output from  clusters changes during their evolution as well as 
its dependence on metallicity. Our goal is to establish the  cluster age and metallicity that are most 
favorable for the production of \heii\ ionizing photons. These predictions 
could be compared with observations of \heii\ emitting galaxies and their star- and cluster-formation 
histories. }

To estimate the temperature and the luminosity of a model star cluster wind,
we used the \citet{CC85} hydrodynamic  solutions coupled with  
stellar feedback parameters -- kinetic energy input ($\dot{E}$) and mass input ($\dot{M}$) --  
prescribed by the {\em Starburst99} model \citep{SB99}. 
 For clarity, we assumed an instantaneous starburst. According to the formalism developed in \citet{Osk2005}, 
 within the mass and energy production
region of radius $R\sim 200$\,pc, the emission measure and the temperature of hot gas is 
$E\!M(t)\,\propto  \dot{M}^3(\alpha\dot{E})^{-1}$
 and
$T(t)\propto \alpha\dot{E}\dot{M}^{-1}$.  The thermalization efficiency parameter, $\alpha$, was introduced
to account for various
uncertainties on energy balance \citep{CC85, Stevens2003}.

While simple, this  approach allows us to investigate in a controlled manner how 
X-ray properties of a star cluster depend on its age, mass, and  metallicity\footnote{After this 
manuscript was accepted, we became aware of the new models of 
X-ray emission (in the 0.5--2.0\,keV band) from hot star cluster winds by \citet{Franeck2022}.}. 
There are, however, some important shortcomings. The \citet{CC85} solutions are highly idealized 
and do not  account for such important physical processes as\ turbulence, ISM inhomogeneity, 
or thermal conduction. 
The choice of the parameter $\alpha$  is guided by the detailed hydrodynamic models of superbubbles.
\citet{Gray2019} show that outflows generated by massive star 
clusters can undergo catastrophic cooling that suppresses adiabatic superwinds, and they demonstrate that 
such models produce stronger nebular line emission compared to adiabatic outflows.
Recently, \citet{Lancaster2021} investigated the evolution of wind-blown bubbles in a turbulent 
ISM. Their three-dimensional hydrodynamic simulations show that turbulent mixing at the 
bubble-shell interface leads to highly efficient cooling, by which the vast majority of the input 
wind energy is radiated  away. Only $\sim 1-10$\,\%\ of the energy input rate from the wind 
remains in the bubble, and the fraction of remaining energy decreases with time. Taking 
the \citet{Lancaster2021} model as a guidance, we set  $\alpha=0.05$, the average fraction 
of energy input rate that is retained. 

Furthermore, we introduced the parameter $\dot{M}_{\rm cool}$ to account for uncertainties associated with  
mass loading, mixing, and evaporation  such that the total rate of mass input  is  
$\dot{M}=\dot{M_\ast}+\dot{M}_{\rm cool}$ [$M_\odot\,{\rm yr}^{-1}$], where $\dot{M_\ast}$ is the rate of 
mass influx provided by stellar winds and SNe as retrieved from  {\em Starburst99}.
\citet{El-Badry2019} investigate the evaporation of gas into the 
bubble interior. They find that the evaporation increases the density and decreases the temperature by more 
than an order of magnitude. 
They conclude that the hot gas mass, momentum, and energy 
are set by the ambient interstellar density and the efficiency of nonlinear mixing at the bubble-shell 
interface.  In their recent hydrodynamical models, \citet{Lancaster2021} use an input  mass-loss rate per unit stellar mass that is five times higher
than predicted by  {\em Starburst99}. We follow a similar 
approach but,
since $\dot{M}_{\rm cool}$ is observationally poorly constrained,  we chose a conservative estimate: 
in order to roughly account for the mass loading, we set  $\dot{M}_{\rm cool}=\dot{M_\ast}$ (i.e.\ double the predictions of 
 {\em Starburst99}). 


{\rev The \citet{CC85}  solutions  assume an adiabatic, stationary outflow and neglect the  star cluster 
gravity. Therefore, the cluster mass enters only indirectly, via  stellar feedback in the form of combined mass 
and energy deposited by stellar winds and SNe.  
As a star cluster evolves, so does its feedback.  The evolution of the cluster wind's 
X-ray luminosity and temperature  at different metallicities  
is  shown in Fig.\,\ref{fig:txlx}}.  Interestingly, prior to the onset of SNe 
at $\sim 10$\,Myr,  the temperature is lower at lower $Z$.  To understand this, we recall that, 
according to our model,  $T(Z)\propto \dot{E}(Z)/\dot{M}(Z)$, where both the 
mechanic energy, $\dot{E}(Z)$, and the mass input, $\dot{M}(Z)$,  rates are  retrieved from  {\em Starburst99}.  
At the epoch when the mechanic energy input is dominated by stellar winds,  
$\dot{E}\propto \varv_{\rm wind}^2$\change, with the average wind velocity ($\varv_{\rm wind}$) decreasing toward low $Z$ \citep{Leitherer1992}.
Hence, to first order, $T(Z)\propto \varv^2_{\rm wind}(Z)$.  At later times ($> 10$\,Myr), when  
SNe overtake stellar winds in supplying the mass and energy,  the temperature increases 
and becomes independent of $Z$.

Our simulations predict that when the model cluster  becomes $\sim 3$\,Myr old, its associated X-ray 
output sharply increases. The characteristic time during which the X-ray luminosity  is at its highest 
lasts for $t_{\rm c}\sim 10$\,Myr 
(upper panel in Fig.\,\ref{fig:txlx}). {\rev  During this epoch,  a cluster with $M_{\rm cl}=10^6\,M_\odot$ 
powers $L^{\rm cl}_{\rm X}\sim 10^{38.5}$\,\ergs.   Then, assuming that the energy and  mass inputs 
from massive star clusters to the ISM are self-similar and scale linearly with  cluster mass,
\begin{equation}
\log (L^{\rm cl}_{\rm X} / M_{\rm cl} )=32.5 ~~~~ [{\rm erg}\,{\rm s}^{-1}\,M_\odot^{-1}].
\label{eq:lc}
\end{equation}  
}

\section{Diffuse X-ray emission from an ensemble of star clusters in a star-forming galaxy}
\label{sec:gal}

In our model, the X-ray luminosity of hot gas in star-forming galaxies is powered by young 
massive clusters. In this section we consider whether this model can reproduce 
the correlation between  X-ray luminosity and SFR that  is well established empirically  (Eq.\,\ref{eq:sfr40}).

The total specific energy emitted in X-rays by a model star cluster with an age 
between $\sim 3$\,Myr and $\sim 13$\,Myr over the characteristic time $t_c=10$\,Myr  
and per unit mass is 
\begin{equation}
 \log \left(L^{\rm cl}_{\rm X}/M_{\rm cl} \times t_{\rm c}\right) \equiv \log{{\cal T}_{\rm c}} \sim
 47\,[{\rm erg}\,M_\odot^{-1}].
 \label{eq:7}
 \end{equation}
These order of magnitude estimates are confirmed by the numeric integration over time of the 
$L^{\rm cl}_{\rm X}({\rm Time})$ curves shown in the upper panel of Fig.\,\ref{fig:txlx}: 
$\log{{\cal T}_{\rm c}(Z_\odot)} = 47.40$,  
$\log{{\cal T}_{\rm c}(0.2Z_\odot)}= 47.43$, and  
$\log{{\cal T}_{\rm c}(0.05Z_\odot)}= 47.62$.

The cluster-formation 
efficiency, $\Gamma$, is a measurement of the total stellar mass forming in bound star clusters 
with respect to the total stellar mass forming in the galaxy \citep[][and references therein]{Adamo2020}, such that 
the cluster-formation rate  is
\begin{equation}
{\rm CFR}\equiv \Gamma\times {\rm SFR}~~~~ [\msunyr].
\label{eq:cfr}
\end{equation}
According to Fig.\,18 in \citet{Adamo2020}, $\Gamma\sim 0.1\,-\,1$, while on average  $\Gamma\sim 0.3$.  
{\refrep As a next step, we estimated the total X-ray luminosity of diffuse gas in a galaxy heated by 
cluster feedback for a given episode of star formation. Combining Eqs.\,(\ref{eq:7}) and (\ref{eq:cfr}), we have }
\begin{equation}
L_{\rm X}= 3.17\times 10^{-8}\times  \Gamma \times {\rm SFR}\times  {\cal{T}}_{\rm c}  ~~~~~[{\rm erg\,s}^{-1}]. 
\label{eq:psi}
\end{equation}
Using $\log{{\cal T}_{\rm c}}= 47.6$ 
{\change and a cluster efficiency $\Gamma=0.3,$}
this yields 
\begin{equation}
\log{L_{\rm X}} = \log{\rm SFR}  +  \log{\cal{T}_{\rm c}} + \log{\Gamma} -7.5 \approx \log{\rm SFR} + 39.6, 
\label{eq:396}
\end{equation}
thus recovering {\change within $\sim 0.2$ dex} the empirical correlation between the X-ray luminosity of 
hot diffuse gas and the SFR in galaxies (Eq.~\ref{eq:sfr40}). Correspondingly, 
according to Eq.~(\ref{eq:ii1}), the ratio between  \heii\,$\lambda 4686$ and H$\beta$ line 
intensities becomes
\begin{equation}
\log \left(I(4686)/I({\rm H}\beta) \right) \approx \log{q} - 13.4,
\label{eq:lgI} 
\end{equation}
in fair agreement with observations (see Fig.\,{\ref{fig:qT} and Sect.~\ref{sec:obs}).

\section{Comparison with observations of \zw}
\label{sec:zw}

\zw\ is one of the nearest ($d_{\rm I\,Zw\,18}=18.2$\,Mpc) and  most metal-poor ($Z_{\rm I\,Zw\,18}\approx 0.02Z_\odot$) 
dwarf galaxies.
{\change It contains two large \hii\ regions ionized by star clusters with ages between
a few to a hundred megayears  \citep{Hunter1995, Papaderos2002, Contreras2011, Rickards2021}. }
{\refrep
\zw\ is filled with diffuse \heii\ emission. From the observed luminosity in the \heii\,$\lambda 4686$ line, 
\citet{Kehrig2015} estimate a rate of \heii\ ionizing photons as 
$\log{Q({\rm He}^+)}\approx  50$\,[s$^{-1}$]. 
}

\citet{Martin1996} proposed that a superbubble is responsible for  the ISM structure in \zw.  
Assuming that the superbubble solely produces the  \heii\ ionizing flux, and  invoking Eq.\,(\ref{eq:q}),  
$\log{Q({\rm He}^+)}=\log{q}+\log{L_{\rm X}}\approx  50$.  To assess whether this is realistic, 
we estimated the specific ionizing parameter, $q$. 
\citet{Annibali2013} show that the main body of \zw\  has 
been forming stars very actively over the past $\sim 10$ Myr.   The bulk of star clusters in \zw\ are therefore younger than 
10-20 Myr.  
For clusters of such ages and metallicities,   the temperature 
of  hot plasma 
filling a superbubble is  $\sim 0.1$\,keV ($1$\,MK),  implying $\log{q}\sim 11$   (Figs.\,\ref{fig:qT} and \ref{fig:txlx}). 
Correspondingly,  
 $\log{L_{\rm X}} \sim 39$\,[\ergs]  is sufficient to provide the required \heii\ ionizing flux. 

\begin{figure}[]
\centering
\includegraphics[width=0.99\columnwidth]{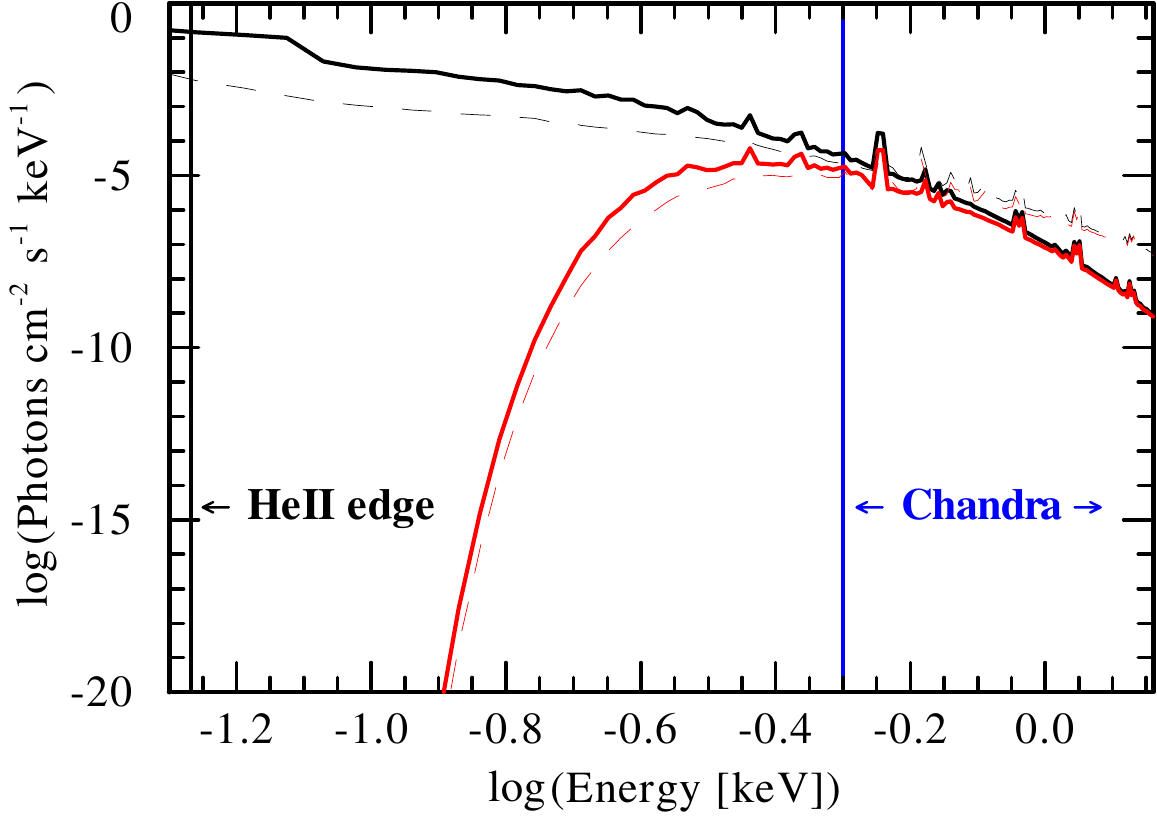}
\caption{Model spectra of diffuse X-ray emission in \zw. The plasma model is by {\sc apec,} with 
$Z=0.02Z_\odot$. 
The thick solid red line represents the model with $kT=0.1$\,keV and X-ray flux 
$F_{\rm X}=2\times 10^{-15}$\,\,erg\,s$^{-1}$\,cm$^{-2}$ in the
0.5--10.0\,keV band (marked as {\em Chandra} in the plot), i.e.,\ the same as reported by \citet{Thuan2004} 
for the extended X-ray emission. The ISM absorption model component ({\em tbabs}) corresponding to the 
foreground absorption toward \zw,  $N_{\rm H}=1.3\times 10^{21}$\,cm$^{-2}$, is applied. 
The thin dashed red line has the same meaning as the solid red line but for $kT=0.2$\,keV.
The thick solid black line has the same meaning as the solid red line but without accounting for the ISM absorbing column 
toward \zw. The X-ray luminosity in the 0.3-10\,keV band is $L_{\rm X}=10^{39}$\,erg\,s$^{-1}$.
The thin dashed black line has the same meaning as the solid black line but for $kT=0.2$\,keV.  }
\label{fig:xzw}
\end{figure}

\citet{Annibali2013} derive for \zw\ an average  SFR of $\sim 1$\,\msunyr\ over the past $\sim 10$\,Myr  and 
significantly less before that \citep[we note that other methods find generally lower values,  e.g.,][]{Leb2017}. 
Adopting  SFR\,$\sim 1$\,\msunyr,   $\log{L_{\rm X}} \sim 39$\,[\ergs], which is\  broadly consistent 
with the predictions of Eqs.~(\ref{eq:sfr40}) and (\ref{eq:396}).  
These estimates show  that the \heii\ ionization in \zw\ can be explained by stellar 
feedback and the superbubble it produced. 

This mechanism  is consistent with  observations. Recently, \citet{Kehrig2021} 
reviewed up-to-date multiwavelength observations of \zw. The total X-ray luminosity of this galaxy,  
$L_{\rm X}(0.5-8.0\,{\rm keV})\msim 10^{40}$\,\ergs, is dominated by a HMXB associated with 
the most populous northwestern star-forming region. The HMXB experiences state transitions 
over a timescale of years, but so far its luminosity has always exceeded $10^{39}$\,\ergs. While {\em XMM-Newton}
yields excellent spectra of the HMXB, only {\em Chandra} can resolve the point and diffuse X-ray 
sources in \zw. 

Indeed, \citet{Thuan2004} reported  the discovery of extended X-ray emission in the main body of \zw. 
Since only $\approx 23$ counts from this diffuse emission were measured,  its spectrum cannot be well constrained.   
To determine the X-ray flux, \citet{Thuan2004} adopted the neutral hydrogen column density of 
$N_{\rm H} =1.3\times 10^{21}$\,cm$^{-2}$ and a thermal plasma model with 11.6\,MK 
(see their Table\,1). This results in  $F_{\rm X}(0.5-10.0\,{\rm keV})\approx 2\times 10^{-15}$\,erg\,s$^{-1}$\,cm$^{-2}$
and  $L_{\rm X}(0.5-10.0\,{\rm keV}) \approx 5\times 10^{37}$\,\ergs.   

However, according to our model the plasma temperature in \zw\ is significantly lower. 
Using the {\sc  apec} model
with 
1.16\,MK and $Z=0.02\,Z_\odot$,  the absorption-corrected X-ray luminosity of 
diffuse gas in \zw\ is 
$L_{\rm X}^{\rm I\,Zw\,18}(0.3-10\,{\rm keV})\approx 10^{39}$\,erg\,s$^{-1}$, which is\, in agreement with our model 
predictions.

The extended X-ray radiation detected by {\em Chandra} is likely only a small fraction of the warm-hot gas filling this galaxy.
At the distance of \zw,  the X-ray luminosity $\log{L_{\rm X}} \sim 39$\,[\ergs] 
corresponds to the unabsorbed flux  $F_{\rm X}=2.5\times 10^{-14}$\,erg\,cm$^{-2}$\,s$^{-1}$. 
However, the relatively high column density severely hampers the ability to detect soft X-rays
(Fig.\,\ref{fig:xzw}). To check whether the existing X-ray observations are suitable for detecting the 
soft emission predicted by our model, we employed the current version of the tool `PIMMS v4.11a: with 
ACIS Pile up and Background Count Estimation' provided by the {\em Chandra} X-Ray Center\footnote{\url{https://cxc.harvard.edu/toolkit/pimms.jsp}}. Unfortunately,  {\em Chandra} 
Cycle\,1 is not included in this version; therefore, we made estimates for the earliest available, 
Cycle\,3. We assumed that the size of the superbubble is similar to the size of the central \heii\ region (i.e.,\
$600$\,\arcsec$^2$). The total background count rate of the ACIS-S detector in the the $600$\arcsec$^2$ area is 
$5\times 10^{-4}$\,s$^{-1}$, which exceeds the count rate expected from the 1\,MK degree superbubble in \zw. 
Therefore, it is clear why the bulk of soft X-ray emission ionizing \heii\ in \zw\ could not be detected 
by {\em Chandra}.}

Importantly, radio observations confirm the presence of a superbubble in \zw. 
The spatial distribution of the radio lobes, the extended X-ray emission, and the \heii\ regions 
are remarkably similar \citep{Hunt2005, Rickards2021,Kehrig2021}.
{\refrep We conclude that the feedback from a few clusters that are $<10$\,Myr old could 
power superbubbles that emit radiation capable of sustaining the bulk of \heii\ nebula emission 
in \zw. Future combined hydrodynamic and radiative transfer models are needed to simulate
the nebula \heii\ emission maps and compare them to the observations.} 

\section{Discussion and summary}
\label{sec:sum}

\begin{table}
\caption{\heii\ ionizing  rate in different types of objects 
coexisting in a superstar cluster older than $\sim 5$\,Myr. 
The metallicity  $Z=0.07$ \zsun\ is adopted. 
Numbers are given for an X-ray luminosity $L_{\rm X}(0.3-10.0\,{\rm keV})=10^{40}$\,erg\,s$^{-1}$
of the superbubble (SBB).}
\begin{center}
\begin{tabular}{llcc}
\hline \hline
Object & Model$^{\ast}$ & Temperature & $\log Q({\rm He\,II})$ \\
            &                          &                      &     [s$^{-1}$              ]\\
\hline \hline
WC star  &  WC 19-18  & $T_\ast=200$\,kK &  48.5 \\
WNE star &  WNE 18-13 & $T_\ast=178$\,kK &  48.4 \\
%
SBB & {\em apec}  & $kT=0.5$\,keV  & 49.7 \\ 
SBB & {\em apec}  & $kT=0.2$\,keV  & 50.2 \\
SBB & {\em apec}  & $kT=0.1$\,keV  & 51.0 \\ \hline \hline
\end{tabular}
\end{center}
{\small (*) For WR stars, $T_\ast$ is the stellar temperature, and the model number is according 
to the model grid computed with the Potsdam Wolf-Rayet (PoWR) model atmsophere code 
accessible at {\url{www.astro.physik.uni-potsdam.de/PoWR}}.
}
\label{tab:qwr}
\end{table}

We propose a  model that explains nebular \heii\ emission in star-forming 
galaxies with ionizing radiation produced by superbubbles. 
We show that when the plasma temperature is sufficiently low 
($kT \la 0.2$\,keV), the specific \heii\ ionizing flux $\log{q}> 10$, which is\ sufficiently 
high to reproduce the ratio between  \heii\,$\lambda 4686$ and H$\beta$ line intensities  
observed in star-forming galaxies.

At temperatures of a few megakelvins, which are typical for hot superbubbles, helium is collisionally ionized. According to the 
calculations of ionization balance in the collisional ionization equilibrium gas,  at temperatures above 1\,MK 
all helium is in the H\,{\sc iii} ionization stage  \citep[the fraction of H\,{\sc ii}  is  
$>5$ orders of magnitude lower than the fraction of He\,{\sc iii};][]{SD1993}. Therefore, strong emission in the 
\heii\,$\lambda 4686$ line within the bubbles is not  expected. 


In order to gain a physical understanding, we developed a simple model 
based on the hydrodynamic solutions for 
cluster winds coupled with the output of {\em Starburst99} models (see Sect.~\ref{sec:sb}).
We find that slower stellar winds at lower $Z$ lead to lower energy input,  
and consequently lower temperatures of hot gas, and result in a higher output of \heii\ 
ionizing photons (Figs.\,\ref{fig:qT}  and  \ref{fig:txlx}).   

To estimate the X-ray luminosities of cluster winds and associated superbubbles, we made two strong
assumptions on cooling and mass loading. For quantitative estimates, we adopted the 
low thermalization efficiency parameter $\alpha=0.05$ and doubled the rate of mass input 
compared to the predictions of {\it Starburst99} models. 
As discussed in \citet{Lancaster2021}, the stronger cooling is favored 
in the regions with higher ambient density. It remains to be seen whether these conditions are 
fulfilled  in galaxies that emit nebular \heii\ spectra. 

In order to estimate the \heii\ ionizing power of the whole ensemble of young star clusters in 
a star-forming galaxy,  we  assumed that superbubbles are self-similar and that their X-ray luminosities are proportional to the mass of the cluster
(we only considered massive clusters capable of driving  
winds when a few megayears old). 
We show that the X-ray luminosity of hot diffuse gas in a galaxy is proportional to its
SFR as $\log{L_{\rm X}}\approx \log{\rm SFR} + 39.6$ \ergs, similar to observations. 
According to our model, the observed prevalence of \heii\ nebulae in low-$Z$ galaxies is 
due to the combination of (1) lower temperatures of superbubbles
and (2) the linear scaling between the X-ray output star clusters and the SFR of their host galaxy 
(Eqs.\,\ref{eq:lc}--\ref{eq:lgI}).  
To test our model, we considered the template 
low-$Z$ galaxy \zw\ and find that \heii\ ionization in this galaxy could be powered by a hot 
superbubble.

{\refrep The approach we follow in this paper is simple -- we used a single-temperature hot 
bubble as a source of ionizing photons and did not consider  the effects of radiative transfer. 
The gas temperature determines the SED of radiation emitted by the bubble. In realistic 
superbubbles the plasma is multi-temperature and could be out of collisional equilibrium \citep{Kavanagh2020}. 
These effects should be included in future detailed modeling.  The SED of the radiation source 
does not only matter for estimating the direct number of \heii\ ionizing photons it produces: the energies of photoelectrons also depend on the SED, and secondary electrons may also 
contribute to the ionization \citep{Valdes2008}.  These effects are, however,  difficult to model for \heii. 

Furthermore, in star-forming galaxies the ISM has a complex 
distribution. This could have a non-negligible effect on the propagation of 
ionizing photons  \citep{Eide2018, Glatzle2019}. 
The final predictions of 
line intensities might be significantly affected by  radiative transfer  
through gas and dust. Hence, computing the intrinsic ionizing flux is only the 
first step of a process investigating radiative feedback. Further work on
radiative transfer simulations is required to connect the intrinsic X-rays fluxes 
with the  observed \heii\ nebula lines. }

Superbubbles are not the only objects that produce \heii\ ionizing flux in star-forming galaxies. 
The emission of various objects that coexist in a superstar cluster that is a few megayears old into the He$^+$ ionizing flux is presented in Table\,\ref{tab:qwr}.
The \heii\ ionizing power of the hottest WR stars is orders of magnitude lower than that of a superbubble. 
{\change To produce numbers of \heii\ ionizing photons comparable to that produced by a superbubble with $ kT=0.2$ keV
and $L_X =10^{40}$\,\ergs, at least $\sim 50$ very hot WR stars would be needed.
Alternatively, a single ULX with $L_X \sim 10^{40}$ \ergs\ and a sufficiently soft spectrum 
could also power the observed \heii\ recombination lines \citep[see][]{Simmonds2021}. }

To conclude, we make the falsifiable prediction that the temperature in superbubbles around  
2--10\,Myr old superstar clusters is lower in lower-$Z$ galaxies. 
The model we suggest here paves the way for stellar feedback  simulations that account consistently for 
evolution, hydrodynamics, and photoionization in star-forming galaxies  that exhibit \heii\ nebular emission.

\begin{acknowledgements}
Authors are grateful to the anonymous referee for the useful report that helped to improve the paper.  
\end{acknowledgements}


\begin{thebibliography}{}
\expandafter\ifx\csname natexlab\endcsname\relax\def\natexlab#1{#1}\fi

\bibitem[{{Adamo} {et~al.}(2020){Adamo}, {Zeidler}, {Kruijssen}, {Chevance},
  {Gieles}, {Calzetti}, {Charbonnel}, {Zinnecker}, \& {Krause}}]{Adamo2020}
{Adamo}, A., {Zeidler}, P., {Kruijssen}, J.~M.~D., {et~al.} 2020, \ssr, 216, 69

\bibitem[{{Annibali} {et~al.}(2013){Annibali}, {Cignoni}, {Tosi}, {van der
  Marel}, {Aloisi}, {Clementini}, {Contreras Ramos}, {Fiorentino}, {Marconi},
  \& {Musella}}]{Annibali2013}
{Annibali}, F., {Cignoni}, M., {Tosi}, M., {et~al.} 2013, \aj, 146, 144

\bibitem[{{Arnaud}(1996)}]{xspec1996}
{Arnaud}, K.~A. 1996, in Astronomical Society of the Pacific Conference Series,
  Vol. 101, Astronomical Data Analysis Software and Systems V, ed. G.~H.
  {Jacoby} \& J.~{Barnes}, 17

\bibitem[{{Barrow}(2019)}]{Barrow2019}
{Barrow}, K. S.~S. 2019, \mnras, 2947

\bibitem[{{Binder} {et~al.}(2018){Binder}, {Levesque}, \&
  {Dorn-Wallenstein}}]{Binder2018}
{Binder}, B., {Levesque}, E.~M., \& {Dorn-Wallenstein}, T. 2018, \apj, 863, 141

\bibitem[{{Cassata} {et~al.}(2013){Cassata}, {Le F{\`e}vre}, {Charlot},
  {Contini}, {Cucciati}, {Garilli}, {Zamorani}, {Adami}, {Bardelli}, {Le Brun},
  {Lemaux}, {Maccagni}, {Pollo}, {Pozzetti}, {Tresse}, {Vergani}, {Zanichelli},
  \& {Zucca}}]{Cassata2013}
{Cassata}, P., {Le F{\`e}vre}, O., {Charlot}, S., {et~al.} 2013, \aap, 556, A68

\bibitem[{{Cheng} {et~al.}(2021){Cheng}, {Wang}, \& {Lim}}]{Cheng2021}
{Cheng}, Y., {Wang}, Q.~D., \& {Lim}, S. 2021, \mnras, 504, 1627

\bibitem[{{Chevalier} \& {Clegg}(1985)}]{CC85}
{Chevalier}, R.~A. \& {Clegg}, A.~W. 1985, \nat, 317, 44

\bibitem[{{Contreras Ramos} {et~al.}(2011){Contreras Ramos}, {Annibali},
  {Fiorentino}, {Tosi}, {Aloisi}, {Clementini}, {Marconi}, {Musella}, {Saha},
  \& {van der Marel}}]{Contreras2011}
{Contreras Ramos}, R., {Annibali}, F., {Fiorentino}, G., {et~al.} 2011, \apj,
  739, 74

\bibitem[{{Dionne} \& {Robert}(2006)}]{Dionne2006}
{Dionne}, D. \& {Robert}, C. 2006, \apj, 641, 252

\bibitem[{{Doughty} \& {Finlator}(2021)}]{Doughty2021}
{Doughty}, C. \& {Finlator}, K. 2021, \mnras, 505, 2207

\bibitem[{{Eide} {et~al.}(2018){Eide}, {Graziani}, {Ciardi}, {Feng},
  {Kakiichi}, \& {Di Matteo}}]{Eide2018}
{Eide}, M.~B., {Graziani}, L., {Ciardi}, B., {et~al.} 2018, \mnras, 476, 1174

\bibitem[{{El-Badry} {et~al.}(2019){El-Badry}, {Ostriker}, {Kim}, {Quataert},
  \& {Weisz}}]{El-Badry2019}
{El-Badry}, K., {Ostriker}, E.~C., {Kim}, C.-G., {Quataert}, E., \& {Weisz},
  D.~R. 2019, \mnras, 490, 1961

\bibitem[{{Franeck} {et~al.}(2022){Franeck}, {W{\"u}nsch},
  {Mart{\'\i}nez-Gonz{\'a}lez}, {Orlitov{\'a}}, {Boorman}, {Svoboda},
  {Sz{\'e}csi}, \& {Douna}}]{Franeck2022}
{Franeck}, A., {W{\"u}nsch}, R., {Mart{\'\i}nez-Gonz{\'a}lez}, S., {et~al.}
  2022, arXiv e-prints, arXiv:2201.12339

\bibitem[{{Glatzle} {et~al.}(2019){Glatzle}, {Ciardi}, \&
  {Graziani}}]{Glatzle2019}
{Glatzle}, M., {Ciardi}, B., \& {Graziani}, L. 2019, \mnras, 482, 321

\bibitem[{{G{\"o}tberg} {et~al.}(2019){G{\"o}tberg}, {de Mink}, {Groh},
  {Leitherer}, \& {Norman}}]{Gotberg2019}
{G{\"o}tberg}, Y., {de Mink}, S.~E., {Groh}, J.~H., {Leitherer}, C., \&
  {Norman}, C. 2019, \aap, 629, A134

\bibitem[{{Gray} {et~al.}(2019){Gray}, {Oey}, {Silich}, \&
  {Scannapieco}}]{Gray2019}
{Gray}, W.~J., {Oey}, M.~S., {Silich}, S., \& {Scannapieco}, E. 2019, \apj,
  887, 161

\bibitem[{{Guseva} {et~al.}(2000){Guseva}, {Izotov}, \& {Thuan}}]{Guseva2000}
{Guseva}, N.~G., {Izotov}, Y.~I., \& {Thuan}, T.~X. 2000, \apj, 531, 776

\bibitem[{{Hunt} {et~al.}(2005){Hunt}, {Dyer}, \& {Thuan}}]{Hunt2005}
{Hunt}, L.~K., {Dyer}, K.~K., \& {Thuan}, T.~X. 2005, \aap, 436, 837

\bibitem[{{Hunter} \& {Thronson}(1995)}]{Hunter1995}
{Hunter}, D.~A. \& {Thronson}, Harley~A., J. 1995, \apj, 452, 238

\bibitem[{{Izotov} {et~al.}(2012){Izotov}, {Thuan}, \& {Privon}}]{Izotov2012}
{Izotov}, Y.~I., {Thuan}, T.~X., \& {Privon}, G. 2012, \mnras, 427, 1229

\bibitem[{{Kavanagh} {et~al.}(2020){Kavanagh}, {Sasaki}, {Breitschwerdt}, {de
  Avillez}, {Filipovi{\'c}}, {Galvin}, {Haberl}, {Hatzidimitriou}, {Henze},
  {Plucinsky}, {Saeedi}, {Sokolovsky}, \& {Williams}}]{Kavanagh2020}
{Kavanagh}, P.~J., {Sasaki}, M., {Breitschwerdt}, D., {et~al.} 2020, \aap, 637,
  A12

\bibitem[{{Kehrig} {et~al.}(2021){Kehrig}, {Guerrero}, {V{\'\i}lchez}, \&
  {Ramos-Larios}}]{Kehrig2021}
{Kehrig}, C., {Guerrero}, M.~A., {V{\'\i}lchez}, J.~M., \& {Ramos-Larios}, G.
  2021, \apjl, 908, L54

\bibitem[{{Kehrig} {et~al.}(2015){Kehrig}, {V{\'\i}lchez}, {P{\'e}rez-Montero},
  {Iglesias-P{\'a}ramo}, {Brinchmann}, {Kunth}, {Durret}, \&
  {Bayo}}]{Kehrig2015}
{Kehrig}, C., {V{\'\i}lchez}, J.~M., {P{\'e}rez-Montero}, E., {et~al.} 2015,
  \apjl, 801, L28

\bibitem[{{Keller} {et~al.}(2014){Keller}, {Wadsley}, {Benincasa}, \&
  {Couchman}}]{Keller2014}
{Keller}, B.~W., {Wadsley}, J., {Benincasa}, S.~M., \& {Couchman}, H.~M.~P.
  2014, \mnras, 442, 3013

\bibitem[{{Kewley} \& {Dopita}(2002)}]{Kewley2002}
{Kewley}, L.~J. \& {Dopita}, M.~A. 2002, \apjs, 142, 35

\bibitem[{{Lancaster} {et~al.}(2021){Lancaster}, {Ostriker}, {Kim}, \&
  {Kim}}]{Lancaster2021}
{Lancaster}, L., {Ostriker}, E.~C., {Kim}, J.-G., \& {Kim}, C.-G. 2021, \apj,
  914, 90

\bibitem[{{Lebouteiller} {et~al.}(2017){Lebouteiller}, {P{\'e}quignot},
  {Cormier}, {Madden}, {Pakull}, {Kunth}, {Galliano}, {Chevance}, {Heap},
  {Lee}, \& {Polles}}]{Leb2017}
{Lebouteiller}, V., {P{\'e}quignot}, D., {Cormier}, D., {et~al.} 2017, \aap,
  602, A45

\bibitem[{{Lehmer} {et~al.}(2021){Lehmer}, {Eufrasio}, {Basu-Zych}, {Doore},
  {Fragos}, {Garofali}, {Kovlakas}, {Williams}, {Zezas}, \&
  {Santana-Silva}}]{Lehmer2021}
{Lehmer}, B.~D., {Eufrasio}, R.~T., {Basu-Zych}, A., {et~al.} 2021, \apj, 907,
  17

\bibitem[{{Leitherer} {et~al.}(1992){Leitherer}, {Robert}, \&
  {Drissen}}]{Leitherer1992}
{Leitherer}, C., {Robert}, C., \& {Drissen}, L. 1992, \apj, 401, 596

\bibitem[{{Leitherer} {et~al.}(1999){Leitherer}, {Schaerer}, {Goldader},
  {Delgado}, {Robert}, {Kune}, {de Mello}, {Devost}, \& {Heckman}}]{SB99}
{Leitherer}, C., {Schaerer}, D., {Goldader}, J.~D., {et~al.} 1999, \apjs, 123,
  3

\bibitem[{{Mac Low} \& {McCray}(1988)}]{MacLow1988}
{Mac Low}, M.-M. \& {McCray}, R. 1988, \apj, 324, 776

\bibitem[{{Martin}(1996)}]{Martin1996}
{Martin}, C.~L. 1996, \apj, 465, 680

\bibitem[{{Mewe} {et~al.}(1985){Mewe}, {Gronenschild}, \& {van den
  Oord}}]{mekal}
{Mewe}, R., {Gronenschild}, E.~H.~B.~M., \& {van den Oord}, G.~H.~J. 1985,
  \aaps, 62, 197

\bibitem[{{Mineo} {et~al.}(2012){Mineo}, {Gilfanov}, \& {Sunyaev}}]{Mineo2012}
{Mineo}, S., {Gilfanov}, M., \& {Sunyaev}, R. 2012, \mnras, 426, 1870

\bibitem[{{Moon} {et~al.}(2011){Moon}, {Harrison}, {Cenko}, \&
  {Shariff}}]{Moon2011Large-Highly-Io}
{Moon}, D.-S., {Harrison}, F.~A., {Cenko}, S.~B., \& {Shariff}, J.~A. 2011,
  \apjl, 731, L32

\bibitem[{{Oskinova}(2005)}]{Osk2005}
{Oskinova}, L.~M. 2005, \mnras, 361, 679

\bibitem[{{Oskinova} {et~al.}(2019){Oskinova}, {Bik}, {Mas-Hesse}, {Hayes},
  {Adamo}, {{\"O}stlin}, {F{\"u}rst}, \& {Ot{\'\i}-Floranes}}]{Oskinova2019}
{Oskinova}, L.~M., {Bik}, A., {Mas-Hesse}, J.~M., {et~al.} 2019, \aap, 627, A63

\bibitem[{{Pakull} \& {Mirioni}(2002)}]{Pakull2002}
{Pakull}, M.~W. \& {Mirioni}, L. 2002, arXiv e-prints, arXiv:0202488

\bibitem[{{Pakull} \& {Mirioni}(2003)}]{Pakull2003}
{Pakull}, M.~W. \& {Mirioni}, L. 2003, in Revista Mexicana de Astronomia y
  Astrofisica Conference Series, ed. J.~{Arthur} \& W.~J. {Henney}, Vol.~15,
  197--199

\bibitem[{{Papaderos} {et~al.}(2002){Papaderos}, {Izotov}, {Thuan}, {Noeske},
  {Fricke}, {Guseva}, \& {Green}}]{Papaderos2002}
{Papaderos}, P., {Izotov}, Y.~I., {Thuan}, T.~X., {et~al.} 2002, \aap, 393, 461

\bibitem[{{P{\'e}rez-Montero} {et~al.}(2020){P{\'e}rez-Montero}, {Kehrig},
  {V{\'\i}lchez}, {Garc{\'\i}a-Benito}, {Duarte Puertas}, \&
  {Iglesias-P{\'a}ramo}}]{PerezMontero2020}
{P{\'e}rez-Montero}, E., {Kehrig}, C., {V{\'\i}lchez}, J.~M., {et~al.} 2020,
  \aap, 643, A80

\bibitem[{{Plat} {et~al.}(2019){Plat}, {Charlot}, {Bruzual}, {Feltre},
  {Vidal-Garc{\'\i}a}, {Morisset}, {Chevallard}, \& {Todt}}]{Plat_2019}
{Plat}, A., {Charlot}, S., {Bruzual}, G., {et~al.} 2019, \mnras, 490, 978

\bibitem[{{Rickards Vaught} {et~al.}(2021){Rickards Vaught}, {Sandstrom}, \&
  {Hunt}}]{Rickards2021}
{Rickards Vaught}, R.~J., {Sandstrom}, K.~M., \& {Hunt}, L.~K. 2021, \apjl,
  911, L17

\bibitem[{{Sander} \& {Vink}(2020)}]{Sander2020}
{Sander}, A. A.~C. \& {Vink}, J.~S. 2020, \mnras, 499, 873

\bibitem[{{Saxena} {et~al.}(2020){Saxena}, {Pentericci}, {Schaerer},
  {Schneider}, {Amorin}, {Bongiorno}, {Calabr{\`o}}, {Castellano}, {Cimatti},
  {Cullen}, {Fontana}, {Fynbo}, {Hathi}, {McLeod}, {Talia}, \&
  {Zamorani}}]{Saxena2020}
{Saxena}, A., {Pentericci}, L., {Schaerer}, D., {et~al.} 2020, \mnras, 496,
  3796

\bibitem[{{Schaerer}(1996)}]{Schaerer1996}
{Schaerer}, D. 1996, \apjl, 467, L17

\bibitem[{{Schaerer} {et~al.}(2019){Schaerer}, {Fragos}, \&
  {Izotov}}]{Schaerer2019}
{Schaerer}, D., {Fragos}, T., \& {Izotov}, Y.~I. 2019, \aap, 622, L10

\bibitem[{{Senchyna} {et~al.}(2020){Senchyna}, {Stark}, {Mirocha}, {Reines},
  {Charlot}, {Jones}, \& {Mulchaey}}]{Senchyna2020}
{Senchyna}, P., {Stark}, D.~P., {Mirocha}, J., {et~al.} 2020, \mnras, 494, 941

\bibitem[{{Shirazi} \& {Brinchmann}(2012)}]{Shirazi2012}
{Shirazi}, M. \& {Brinchmann}, J. 2012, \mnras, 421, 1043

\bibitem[{{Silich} {et~al.}(2005){Silich}, {Tenorio-Tagle}, \&
  {A{\~n}orve-Zeferino}}]{Silich2005}
{Silich}, S., {Tenorio-Tagle}, G., \& {A{\~n}orve-Zeferino}, G.~A. 2005, \apj,
  635, 1116

\bibitem[{{Simmonds} {et~al.}(2021){Simmonds}, {Schaerer}, \&
  {Verhamme}}]{Simmonds2021}
{Simmonds}, C., {Schaerer}, D., \& {Verhamme}, A. 2021, \aap, 656, A127

\bibitem[{{Smith} {et~al.}(2018){Smith}, {Campbell}, {Struck}, {Soria},
  {Swartz}, {Magno}, {Dunn}, \& {Giroux}}]{Smith2018}
{Smith}, B.~J., {Campbell}, K., {Struck}, C., {et~al.} 2018, \aj, 155, 81

\bibitem[{{Smith} {et~al.}(2019){Smith}, {Wagstaff}, {Struck}, {Soria}, {Dunn},
  {Swartz}, \& {Giroux}}]{Smith2019}
{Smith}, B.~J., {Wagstaff}, P., {Struck}, C., {et~al.} 2019, \aj, 158, 169

\bibitem[{{Smith} {et~al.}(2001){Smith}, {Brickhouse}, {Liedahl}, \&
  {Raymond}}]{apec}
{Smith}, R.~K., {Brickhouse}, N.~S., {Liedahl}, D.~A., \& {Raymond}, J.~C.
  2001, \apjl, 556, L91

\bibitem[{{Stevens} \& {Hartwell}(2003)}]{Stevens2003}
{Stevens}, I.~R. \& {Hartwell}, J.~M. 2003, \mnras, 339, 280

\bibitem[{{Strickland} {et~al.}(2004){Strickland}, {Heckman}, {Colbert},
  {Hoopes}, \& {Weaver}}]{Strickland2004}
{Strickland}, D.~K., {Heckman}, T.~M., {Colbert}, E. J.~M., {Hoopes}, C.~G., \&
  {Weaver}, K.~A. 2004, \apj, 606, 829

\bibitem[{{Sutherland} \& {Dopita}(1993)}]{SD1993}
{Sutherland}, R.~S. \& {Dopita}, M.~A. 1993, \apjs, 88, 253

\bibitem[{{Sz{\'e}csi} {et~al.}(2015){Sz{\'e}csi}, {Langer}, {Yoon}, {Sanyal},
  {de Mink}, {Evans}, \& {Dermine}}]{Szecsi2015}
{Sz{\'e}csi}, D., {Langer}, N., {Yoon}, S.-C., {et~al.} 2015, \aap, 581, A15

\bibitem[{{Thuan} {et~al.}(2004){Thuan}, {Bauer}, {Papaderos}, \&
  {Izotov}}]{Thuan2004}
{Thuan}, T.~X., {Bauer}, F.~E., {Papaderos}, P., \& {Izotov}, Y.~I. 2004, \apj,
  606, 213

\bibitem[{{Thuan} \& {Izotov}(2005)}]{Thuan2005}
{Thuan}, T.~X. \& {Izotov}, Y.~I. 2005, \apjs, 161, 240

\bibitem[{{Vald{\'e}s} \& {Ferrara}(2008)}]{Valdes2008}
{Vald{\'e}s}, M. \& {Ferrara}, A. 2008, \mnras, 387, L8

\end{thebibliography}

\end{document}